\documentclass[12pt,english]{article}
\usepackage[LGR,T1]{fontenc}
\usepackage[latin1]{inputenc}
\usepackage{geometry}
\usepackage{amsmath}
\usepackage{amssymb}
\usepackage{graphicx}
\usepackage{esint}

\makeatletter

\DeclareRobustCommand{\greektext}{%
  \fontencoding{LGR}\selectfont\def\encodingdefault{LGR}}
\DeclareRobustCommand{\textgreek}[1]{\leavevmode{\greektext #1}}
\DeclareFontEncoding{LGR}{}{}
\DeclareTextSymbol{\~}{LGR}{126}
\newcommand{\lyxmathsym}[1]{\ifmmode\begingroup\def\b@ld{bold}
  \text{\ifx\math@version\b@ld\bfseries\fi#1}\endgroup\else#1\fi}

\providecommand{\tabularnewline}{\\}

\@ifundefined{date}{}{\date{}}

\usepackage{units}\usepackage{amstext}\usepackage{dsfont}\usepackage{setspace}\usepackage{mathrsfs}\usepackage{longtable}\usepackage{times}\usepackage{ulem}\usepackage{babel}

\newcommand{\suppress}[1]{}

\newlength\wvtextpercent
\newbox\strikebox
\def\strike#1{\setbox\strikebox \hbox{<#1>}\hbox{\raise0.5ex\hbox to 0pt{\vrule height 0.4pt width \wd\strikebox\hss}\copy\strikebox}}

\makeatother

\usepackage{babel}
\begin{document}


\linespread{1.5}

\title{\textbf{\Huge{}A derivation of a microscopic entropy and time irreversibility
from the discreteness of time}}

\author{\textbf{Roland Riek} \\
 \textbf{Laboratory of Physical Chemistry, ETH Zurich, Switzerland}}

\maketitle
Keywords:

\vspace{3cm}

\linespread{1}

Address for correspondence:\\
 Roland Riek\\
 Laboratory of Physical Chemistry\\
 ETH Zuerich\\
 Wolfgang-Pauli-Strasse 10\\
 HCI F 225\\
 CH-8093 Zurich\\
 Tel.: +41-44-632 61 39\\
 e-mail: roland.riek@phys.chem.ethz.ch

\textbf{\large{}\newpage{} }{\large \par}

\section{Abstract}

All of the basic microsopic physical laws are time reversible. In
contrast, the second law of thermodynamics, which is a macroscopic
physical representation of the world,  is able to describe irreversible
processes in an isolated system through the change of entropy $\triangle S>0$.
It is the attempt of the present manuscript to bridge the microscopic
physical world with its macrosocpic one with an alternative approach
than the statistical mechanics theory of Gibbs and Boltzmann. It is
proposed that time is discrete with constant step size. Its consequence
is the presence of time irreversibility at the microscopic level if
the present force is of complex nature ($F(r)\neq const)$. In order
to compare this discrete time irreversible mechamics (for simplicity
a ``classical'', single particle in a one dimensional space is selected)
with its classical Newton analog, time reversibility is reintroduced
by scaling the time steps for any given time step $n$ by the variable
$s_{n}$ leading to the Nosé-Hoover Lagrangian. The corresponding
Nosé-Hoover Hamiltonian comprises a term $N_{df}\, k_{B\,}T\, ln\, s_{n}$
($k_{B}$ the Boltzmann constant, $T$ the temperature, and $N_{df}$
the number of degrees of freedom) which is defined as the microscopic
entropy$S_{n}$ at time point $n$ multiplied by $T$. Upon ensemble
averaging this microscopic entropy $S_{n}$ in equilibrium for a system
which does not have fast changing forces approximates its macroscopic
counterpart known from thermodynamics. The presented derivation with
the resulting analogy between the ensemble averaged microscopic entropy
and its thermodynamic analog suggests that the original description
of the entropy by Boltzmann and Gibbs is just an ensemble averaging
of the time scaling variable $s_{n}$ which is in equilibrium close
to 1, but that the entropy term itself has its root not in statistical
mechanics but rather in the discreteness of time.

\section{Introduction}

While all the basic (microscopic) physical laws including the fundamental
differential equations of mathematical physics Hamilton\textquoteright{}s,
Lagrange\textquoteright{}s, Maxwell\textquoteright{}s, Newton\textquoteright{}s,
Einstein's, and Schroedinger's are time reversible, only the second
law of thermodynamics describing macroscopic systems brings the arrow
of time into play by requesting that the entropy increases in an isolated
macroscopic system (Landau and Lifshitz, 1980; Prize 1996; Penrose,
2010; Penrose, 1989). Following the systematic formulation of statistical
mechanics by Gibbs and Boltzmann, entropy $S$ reflects thereby the
number of accessible micro-states of the system in study in its thermodynamic
equilibrium (for a micro canonical ensemble $S^{G}=k_{B}\, ln\,\text{\textgreek{W}}$
with $k_{B}$ the Boltzmann constant, and $\text{\textgreek{W}}$
the number of accessible micro-states (Landau and Lifshitz, 1980;
Hoover, 1999, Boltzmann, 1866; Greiner et al., 1993) and exerts its
presence as being part of the total free energy (for example for an
ideal gas the Gibbs free energy is given by $\text{}G\,=U+p\text{}V-T\text{}S$
 with $U$ is the inner Energy, $p$ the pressure, $V$ the volume,
and $T$ the temperature; Landau and Lifshitz, 1980, Hoover, 1999).
The latter argument brings the term entropy $S$ back to its roots,
where Clausius tried to design and understand \textquotedblleft{}heat
engines\textquotedblright{}, which are cyclic machines for the conversion
of heat $Q$ into useful work, and found when averaged over many cycles
that for an irreversible machine $\frac{\text{\textgreek{D}}Q}{T}=\text{\textgreek{D}}S$
> 0 (Clausius, 1865). Although the statistical mechanics argument
of entropy increase with time by describing a system tendency towards
its most probable state, which is the equilibrium state, is sound,
Loschmidt and Zermelo\textquoteright{}s reversal and recurrence objections
remain as powerful as ever (Steckline, 1982; Lohschmidt, 1876, Poincare,
1890). Both objections have been extensively discussed and are just
mentioned here in short. The center of Zermelo argument is that if
waited long enough any system, which started at a special low entropic
state, under a time reversible physics may go back to its special
low entropic starting state requesting a decrease of entropy over
a long period of action, which is in general not observed. Lohschmidt
indicated that any time reversible microscopic process can be reversed
by reversing all the velocity vectors of the involved particles and
if done so yielding a decrease of entropy over a long trajectory,
something that is not observed in general as just mentioned.  

In addition to these objections of the interpretation of entropy,
it is the author\textquoteright{}s notion, that the arrow of time
is such an important physical measure that it has likely its root
deep in physics at the microscopic level and not just at the macroscopic
statistical level \textendash{} a point of view that was also mentioned
beforehand by others (for example Prigogine, 1997; Prize, 1996) and
is experimentally supported by the observed time symmetry violation
of the weak force (Christensen et al., 1964; Lees et al., 2012). The
idea of an arrow of time (at the microscopic level) requests the existence
of time contrasting the view of many scientists and philosophers,
who suggested that time is an illusion (Smolin, 2013). In the following,
we will assume time and the arrow of time to be fundamental already
at the microscopic level.

But from where could the arrow of time, time irreversibility, and
entropy be originated from? As we shall see, it is our attempt to
derive the arrow of time and the entropy part of the total free energy
from the assumption that time is discrete. Although the introduction
of a discrete time is not new and upon energy quantization for quantum
mechanics straightforward (for example Poincare, 1913; Farias and
Recami, 2007; Thomson, 1925) it is not a very popular concept, since
upon the introduction by Newton, time and space are an infinitely
divisible continuum yielding powerful mathematical description in
classical physics, quantum mechanics, as well as special and general
relativity theories. An experimental argument in favor of the discreteness
of time is, that the experimentally measured time is composed of an
array of events, which can be exemplified in physics only by an energy-consuming
clock measurement, and thus can not be measured continuously because
of the uncertainty principle between energy and time ($\text{\textgreek{D}E \textgreek{D}t > h/2}$
with h being the Planck constant).

There have been a few attempts in the literature that developed theories
with time as a discrete parameter (such as Yang, 1947, Levi, 1927;
Caldirola, 1953; Lee et al., 1983; Farias and Erasmo, 2007; Jaroszkiewicz
and Norton, 1997 and 1998; Valsakumar 2005). To describe the time
evolution of Hamiltonian systems with time as a discrete parameter
Lee introduced time as a discrete dynamical variable (Lee, 1983).
Lee points out that throughout the development of physics, time always
appears as a continuous parameter, while the space coordinates in
non-relativistic theories are dynamical operators dependent on time.
Lee showed that the introduction of time as a dynamical variable enables
the discretization of time without violating the conservation of energy
law. Subsequently, by invoking the discrete-time action principle
by Cadzow (1970) Jaroszkiewicz and coworkers (Jaroszkiewicz and Norton,
1997 and 1998) succeeded in developing an equation of motion with
time as a discrete parameter having time steps of equal size. Valsakumar
(2005) took a fresh look at the problem under the assumption that
the time steps are identical to the Planck time ($5.4\,10^{-44}$
s). By adopting the phase space density approach he yields the discrete
time-analog of the Liouville equation of the phase space density in
classical mechanics. Amongst others, his approach yields an arrow
of time that follows from the documentation that the replacement of
the time derivative by a backward difference operator only can preserve
the non-negativity of the phase space density (Valsakumar, 2005).
In parallel, in attempts to unify the general relativity theory with
quantum mechanics including in particular quantum loop theory (Rovelli,
2011; Smolin, 2013) it is assumed at the most fundamental level that
time has a granular structure with the Planck time as the smallest
time step. Furthermore, it appears that the Dirac equation which describes
the free electron is more sound in presence of a discrete time than
its continuous analog without loosing Lorentz invariance (Caldirola,
1953; Farias and Recami, 2007). 

We follow here the approach by Lee (1983) introducing time as a dynamical
discrete variable yielding a scaling of time that depends on the potential
present. This Ansatz is only applied to classical physics because
of simplicity. The consequences of scaling time onto the law of energy
conservation requests a reformulation of the Hamiltonian as established
by Nosé for isothermal molecular dynamics simulations of macroscopic
systems (Nosé, 1984a and 1984b; Nosé, 1986; Hoover, 1985; Hoover,
2007). There is a logarithmic term in the Nosé-Hamiltonian which is
dependent on the scaling of time. As we shall see, this term is defined
as the microscopic entropy and if ensemble-averaged is equivalent
to the macroscopic entropy as also exemplified by the simple example
of the expansion of an ideal gas. After the introduction of the discrete
time (3.1), time irreversibility in a microscopic system is discussed
(3.2), followed by the reintroduction of time reversibility through
scaling of time (3.3) yielding the microscopic entropy of a single
particle (3.5) and a many particles system (3.6). It is then further
shown that the averaged microscopic entropy corresponds to the Boltzmann
entropy (3.7) as well as the Gibbs entropy (3.8 and 3.9). In (3.10)
the microscopic and macroscopic entropies of the volume expanding
gas are calculated, followed by a discussion on time-irreversibility
within time reversible descriptions of physical laws (3.11). In 3.12
an explicit description of the evolution of a single particle within
the established discrete time theory is given, followed by a conclusion.

\newpage{}

\section{Theory}

\subsection{Under a discrete time}

In classical physics, a non-relativistic particle with mass $m$ at
the position $\mathbf{r}(t)$ at a given time $t$ in a potential
$V(\mathbf{r})$ has the following Lagrangian $L=\frac{1}{2}m\mathbf{\dot{r}}^{2}-V(\mathbf{r})$
with $\mathbf{\dot{r}}=\frac{d\mathbf{r}}{dt}$ being the velocity.
The use of bold letters indicate thereby the vector character of the
physical variables such as $\mathbf{\mathbf{r}}$ and $\mathbf{\dot{r}}$$ $.
In examples having a one dimensional space only they are replaced
by normal letters (such as and $\dot{r}$ and $r$). The classical
trajectory of a particle is determined by the minimal extremity of
the action $A_{c}$, which is usually defined by the time integral
over the Lagrangian between the two time points of interest ($t_{i}$
and $t_{f})$:

\begin{equation}
A_{c}=\int_{t_{i}}^{t_{f}}L(\mathbf{r},\mathbf{\dot{r}},t)\, dt
\end{equation}
denoting c for the continuous or classical case. 

In the discrete time formalism, Lee (1983) replaced the continuos
function $\mathbf{r}(t)$ by a sequence of discrete values: 
\begin{equation}
(\mathbf{r}_{0},\, t_{0}),(\mathbf{r}_{1},\, t_{1}),...,(\mathbf{r}_{n},\, t_{n}),...,(\mathbf{r}_{N+1},\, t_{N+1})
\end{equation}
 with $\mathbf{(r}_{0},\, t_{0})$ the initial and $\mathbf{(r}_{N+1},\, t_{N+1})$
the final position. In this description $\mathbf{r}_{n}$ is still
continuous, while $t_{n}$ is discrete. In discrete mechanics there
are many possible definitions of the concomitant velocity $\mathbf{\dot{r}_{n}}$
and acceleration $\mathbf{\ddot{r}}_{n}$. Since it is the attempt
of the present work to introduce a time irreversible microscopic physics,
the velocity at time point $n$ $\mathbf{\dot{r}_{n}}$ is defined
time asymmetric (note, this contrasts to time symmetric definitions
usually used for example in molecular dynamics simulation to obtain
a time-reversible mechanics). Furthermore, it is defined backward
in time in order to determine it from the past permitting a forward
progressing description. This approach is in line with our daily experience
that the presence is determined by the past and presence (such as
experimentally-derived information from the past and presence): 
\begin{equation}
\mathbf{\dot{r}}_{n}\,=\mathbf{\dot{r}_{\mathrm{n}}(r_{\mathrm{n}}\mathrm{,\mathrm{\mathbf{r}_{\mathrm{n-}1};t_{n},t_{n-1})}}}=\frac{\mathbf{r}_{n}-\mathbf{r}_{n-1}}{t_{n}-t_{n-1}}
\end{equation}

Using this description of the velocity the following action in presence
of a discrete time formalism is defined: 
\begin{equation}
A=\sum_{n=1}^{N+1}\left(\frac{1}{2}\frac{m\left(\mathbf{r}_{n}-\mathbf{r}_{n-1}\right)^{2}}{t_{n}-t_{n-1}}-\left(t_{n}-t_{n-1}\right)V(\mathbf{r}_{n})\right)
\end{equation}
(Please note, that the definition of the action is different from
the one defined by Lee (1983) who uses a mean potential to get a symmetric
action along the space coordinate).

According to Lee (1983) the discrete analog of Newton's law can be
derived by setting
\begin{equation}
\frac{\partial A}{\partial\mathbf{r}_{n}}=0
\end{equation}
which yields
\begin{equation}
m(\frac{\mathbf{r}_{n}-\mathbf{r}_{n-1}}{t_{n}-t_{n-1}}-\frac{\mathbf{r}_{n+1}-\mathbf{r}_{n}}{t_{n+1}-t_{n}})-\left(t_{n}-t_{n-1}\right)\frac{\partial V(\mathbf{r}_{n})}{\partial r_{n}}=0
\end{equation}
\begin{equation}
m\,\frac{\frac{\mathbf{r}_{n}-\mathbf{r}_{n-1}}{t_{n}-t_{n-1}}-\frac{\mathbf{r}_{n+1}-\mathbf{r}_{n}}{t_{n+1}-t_{n}}}{t_{n}-t_{n-1}}=\frac{\partial V(\mathbf{r}_{n})}{\partial r_{n}}
\end{equation}
\begin{eqnarray}
\nonumber \\
\nonumber \\
 &  & \boxed{m\,\mathbf{\ddot{r}}_{n}=m\,\frac{\mathbf{\dot{r}}_{n+1}-\mathbf{\dot{r}}_{n}}{t_{n}-t_{n-1}}=-\frac{\partial V(\mathbf{r}_{n})}{\partial\mathbf{r}_{n}}=\mathbf{F}(\mathbf{r}_{n})}
\end{eqnarray}
with $\mathbf{F}(\mathbf{r}_{n})$ is the (vectorial) force of the
potential $V(\mathbf{r}_{n})$ at point $(\mathbf{r}_{n},\, t_{n})$,
which is discrete in time. The acceleration $\mathbf{\ddot{r}}_{n}$
is defined through the Newton's law of eq. 8 by
\begin{equation}
\mathbf{\ddot{r}}_{n}=\mathbf{\ddot{r}}_{n}(\mathbf{F}(\mathbf{r}_{n})\mathrm{,\mathrm{m)}}=\mathbf{\ddot{r}}_{n}(\mathbf{r}_{n+1},\mathbf{r}_{\mathrm{n}}\mathrm{,\mathrm{\mathbf{r}_{\mathrm{n-}1};t_{n+1},t_{n},t_{n-1})}}=\mathbf{\ddot{r}}_{n}(\mathbf{\dot{r}}_{n+1},\mathbf{\dot{r}}_{n\mathrm{}}\mathrm{;t_{n},t_{n-1})}=\frac{\mathbf{\mathbf{\dot{r}}}_{n+1}-\mathbf{\dot{r}}\mathbf{}_{n}}{t_{n}-t_{n-1}}
\end{equation}
With these definitions $\mathbf{\dot{r}}_{n}$ and $\mathbf{\ddot{\mathbf{r}}}_{n}$
are bound to the corresponding time $t_{n}$. $\mathbf{\dot{r}}_{n}$
is defined backward in time, while $\mathbf{\ddot{r}}_{n}$ is defined
by the force $\mathbf{F}(\mathbf{r}_{n})$ through the discrete variant
of the Newton's law of eq. 8. The Newton's law also defines the velocity
forward in time (i.e. $\mathbf{\dot{r}}_{n+1}$) if the time step
size is known (such as $t_{n}-t_{n-1}=const$). These definitions
follow a general correspondence principle between continuous classical
mechanics and the presented discrete mechanics introduced in Table
1 and enables the calculation of a time trajectory. Thus, for any
time series $t_{1},.....t_{N+1},$the positions $\mathbf{r}_{1}.....,\mathbf{r}_{N+1}$,
velocities $\mathbf{\dot{r}}_{1}........\mathbf{\dot{r}}_{N+1}$ and
the acceleration $\mathbf{\ddot{r}}_{1}.......,\mathbf{\ddot{r}}_{N+1}$
can be determined given the starting position $\mathbf{(r}_{0},\, t_{0})$
and the potential $V(\mathbf{r}_{n})$ as discussed in details in
3.12 (note, that $\mathbf{\dot{r}}_{0}$ and $\mathbf{\ddot{r}}_{0}$
are not defined because of the unknown $t_{-1}$, but they are also
not necessary for the determination of the trajectory).

Table 1: Correspondence principle between continuos classical mechanics
and the discrete mechanics presented here

\begin{tabular}{|c|c|}
\hline 
Continuos mechanics & Discrete mechanics (a)\tabularnewline
\hline 
\hline 
$t$ & $t_{n}$\tabularnewline
\hline 
$dt$ & $t_{n}-t_{n-1}$\tabularnewline
\hline 
$\mathbf{r}$ & $\mathbf{r}_{n}$\tabularnewline
\hline 
$\frac{\partial X}{\partial\mathbf{t}}$ & $\frac{\mathbf{\mathrm{X}}_{n}(t_{n})-\mathbf{\mathrm{X}}_{n-1}(t_{n-1})}{t_{n}-t_{n-1}}$
(b)\tabularnewline
\hline 
$\frac{\partial\mathbf{\dot{\mathrm{X}}}}{\partial\mathbf{t}}$ & $\frac{\mathbf{\dot{\mathrm{X}}}_{n+1}(t_{n+1})-\mathbf{\dot{\mathrm{X}}}_{n}(t_{n})}{t_{n}-t_{n-1}}$
(c)\tabularnewline
\hline 
\end{tabular}

(a) other definitions are in principle possible.

(b) $\mathbf{\mathrm{X}}_{n}(t_{n})$ is a time-dependent variable
such as $\mathbf{r}_{n}$ and $L_{n}$.

(c) $\mathbf{\dot{\mathrm{X}}}_{n}(t_{n})$is a time-derivative of
the time-dependent variable $\mathbf{\mathrm{X}}_{n}(t_{n})$ such
as $\mathbf{\dot{r}}_{n}$. 

Alternatively, a Lagrangian Ansatz for the derivation of the Newton
law given a sequence of time points $\{t_{n},\: n\,=\,0,\,1,\,...,\, N+1\}$
in presence of a discrete time can be derived as follows:

We start with the above equation of the Lagrangian (from eq. 4) being
\begin{equation}
L_{n}=L_{n}(\mathbf{r}_{\mathrm{n}}\mathrm{,\mathrm{\mathbf{r}_{\mathrm{n-}1};t_{n},t_{n-1})}}=\frac{1}{2}\frac{m\left(\mathbf{r}_{n}-\mathbf{r}_{n-1}\right)^{2}}{(t_{n}-t_{n-1})^{2}}-V(\mathbf{r_{\mathrm{n}}})=\frac{1}{2}m\mathbf{\dot{r}}_{n}^{2}-V(\mathbf{r_{\mathrm{n}}})=L_{n}(\mathbf{r}_{\mathrm{n}}\mathrm{,\mathrm{\mathbf{\dot{r}}_{\mathrm{n}})}}
\end{equation}
If the discrete analog of the Lagrangian equation ($\frac{\text{}\partial}{\partial t}\frac{\partial L}{\mathbf{\partial\dot{r}_{\mathrm{}}}}=-\frac{\partial L}{\partial\mathbf{r}}$)
is written following the correspondence principle depicted in Table
1
\begin{equation}
\frac{\text{}1}{t_{n}-t_{n-1}}(\frac{\partial L(\mathbf{r}_{\mathrm{n+1}}\mathrm{,\mathrm{\mathbf{\dot{r}}_{\mathrm{n+1}})}}}{\mathbf{\partial\dot{r}_{\mathrm{\mathrm{n+1}}}}}-\frac{\partial L(\mathbf{r}_{\mathrm{n}}\mathrm{,\mathrm{\mathbf{\dot{r}}_{\mathrm{n}})}}}{\partial\mathbf{\dot{r}_{\mathrm{\mathrm{n}}}}})=-\frac{\partial L}{\partial\mathbf{r}_{n}}
\end{equation}
the same discrete Newton' law is obtained as above: $m\,\ddot{\mathbf{r}}_{n}=m\,\frac{\mathbf{\dot{r}}_{n+1}-\mathbf{\dot{r}_{\mathrm{n}}}}{t_{n}-t_{n-1}}=\mathbf{F}(\mathbf{r}_{n})$.

\includegraphics[scale=0.5]{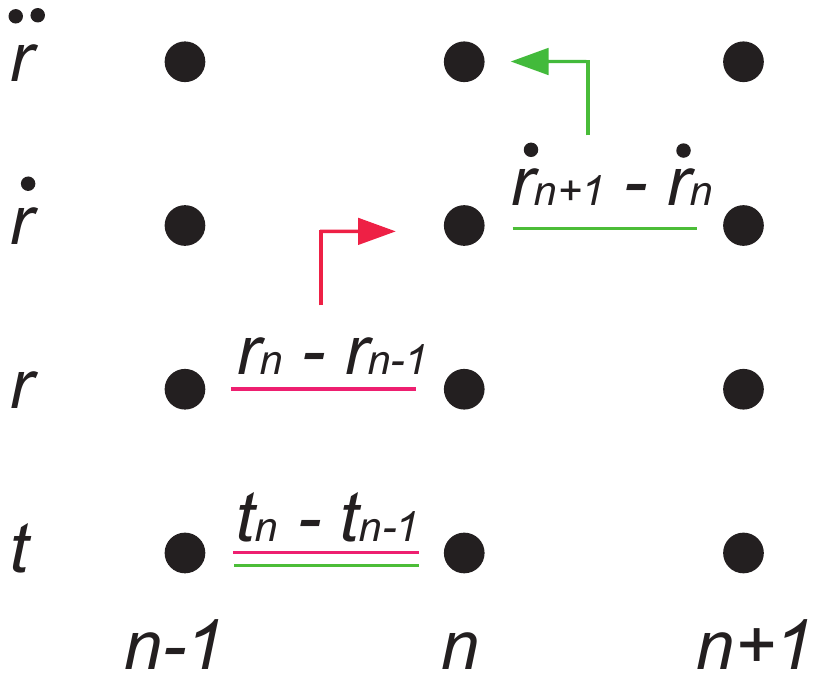}

Figure 1: The dependencies of the velocity $\dot{r}_{n}$ and the
acceleration $\ddot{\mathbf{r}}_{n}$ from the corresponding times
$t_{n}$ and space $r_{n}$ coordinates are graphically indicated.
The discreetness of time is indicated by time points $n-1$, $n$,
and $n+1$. 

\newpage{}

\subsection{Time reversibility/irreversibility in a one dimensional space}

Having established the Newton's law (eq. 8) in presence of a discrete
time, we would like to pursuit the issue of time reversibility or
irreversibility. It is immediately evident from the inspection of
the discrete Newton's law (eq. 8) that under the assumption of having
$\Delta t_{n}=t_{n}-t_{n-1}=const$ the equation is not time symmetric
because of the term $\mathbf{F}(\mathbf{r}_{n})$. Only for $\mathbf{F}(\mathbf{r}_{n})=const$
or/and at the continuous limit for $t_{n}-t_{n-1}\rightarrow0$ with
$lim_{t_{n}->t_{n+1}}\mathbf{F}(\mathbf{r}_{n})=\mathbf{F}(\mathbf{r}_{n+1})$
time reversibility is obtained. This finding is irrespective of whether
the action is symmetric in $\mathbf{r}_{n}$ or not. Hence, the discreteness
of time with a $\Delta t_{n}=const$ in presence of a complex force
(i.e. $\mathbf{F}(\mathbf{r}_{n})\neq const$) yields immediately
time irreversibility and an arrow of time.

More formally, time reversibility can be described by a two step process
having one step forward followed by a step backward. Let us first
consider the evolution of the Newton's law of a single particle with
two step forwards in a one dimensional space (i.e. $\mathbf{r}_{n}$
= $\mathrm{\mathrm{\mathbf{\mathrm{r}}_{\mathrm{n}}}}$ and $\mathbf{F}(\mathrm{\mathrm{\mathrm{\mathbf{r}}}_{n}})=\mathrm{\mathrm{F\mathbf{}}}(r\mathrm{}_{n})=-\frac{\partial V(r_{n})}{\partial r_{n}}$)
\global\long\def\labelenumi{(\roman{enumi})}
 
\begin{enumerate}
\item 
\begin{equation}
\dot{r}_{n+1}=\dot{r}_{n}-\frac{1}{m}(t_{n}-t_{n-1})\frac{\partial V(r_{n})}{\partial r_{n}}
\end{equation}

\item 
\begin{equation}
\dot{r}_{n+2}=\dot{r}_{n+1}-\frac{1}{m}(t_{n+1}-t_{n})\frac{\partial V(r_{n+1})}{\partial r_{n+1}}
\end{equation}

\end{enumerate}
If the second step is now backward in time
\begin{equation}
\dot{r}_{n+2}=\dot{r}_{n+1}+\frac{1}{m}(t_{n+1}-t_{n})\frac{\partial V(r_{n+1})}{\partial r_{n+1}}
\end{equation}
and if time reversibility is requested 
\begin{equation}
\dot{r}_{n}=\dot{r}_{n+2}=\dot{r}_{n+1}+\frac{1}{m}(t_{n+1}-t_{n})\frac{\partial V(r_{n+1})}{\partial r_{n+1}}=\dot{r}_{n}-\frac{1}{m}(t_{n}-t_{n-1})\frac{\partial V(r_{n})}{\partial r_{n}}+\frac{1}{m}(t_{n+1}-t_{n})\frac{\partial V(r_{n+1})}{\partial r_{n+1}}
\end{equation}
Looking at the last equation with $\Delta t_{n}=const$ it is evident
that only with $\frac{\partial V(r_{n})}{\partial r_{n}}=F(r_{n})=const$
eq. 15 is fullfilled and thus the process of interest is reversible.
Under a more complex force however in presence of $\Delta t_{n}=const$
the process is irreversible. That complex processes with many particles
under a complex force are time irreversible is in line with our daily
experiences (such as a cup that falls from a table and breaks into
pieces), and thus sound, albeit most of the physical laws are time
reversible including the Newton's mechanics. It is therefore, the
view of the author that the presented discrete homogeneous time mechanics
(abbreviated diho mechanics) with $\Delta t_{n}=const=\Delta t$ (with
$\Delta t$ probably equal to the Planck time $5.4\,10^{-44}$ s)
may well describe nature, while the continuous, time reversible Newton
mechanics is only an approximation. Usually, this approximation appears
to work very well attributed to the small deviation from time reversibility
because the time interval $\Delta t$ is very short. However, in macroscopic
processes with many particles the Newton mechanics apparently breaks
down since it requests a description of the system by means of thermodynamics
including the entropy term and time irreversibility. If the presented
discrete time mechanics is a more profound theory than the Newton's
mechanics, on the one hand it should reflect irreversible properties
of large complex systems usually described by thermodynamics and statistical
mechanics (i.e. time irreversibility and entropy), while on the other
hand it should under certain boundary conditions asymptotically approximate
the Newton's mechanics and thus time reversibility.

\includegraphics[scale=0.5]{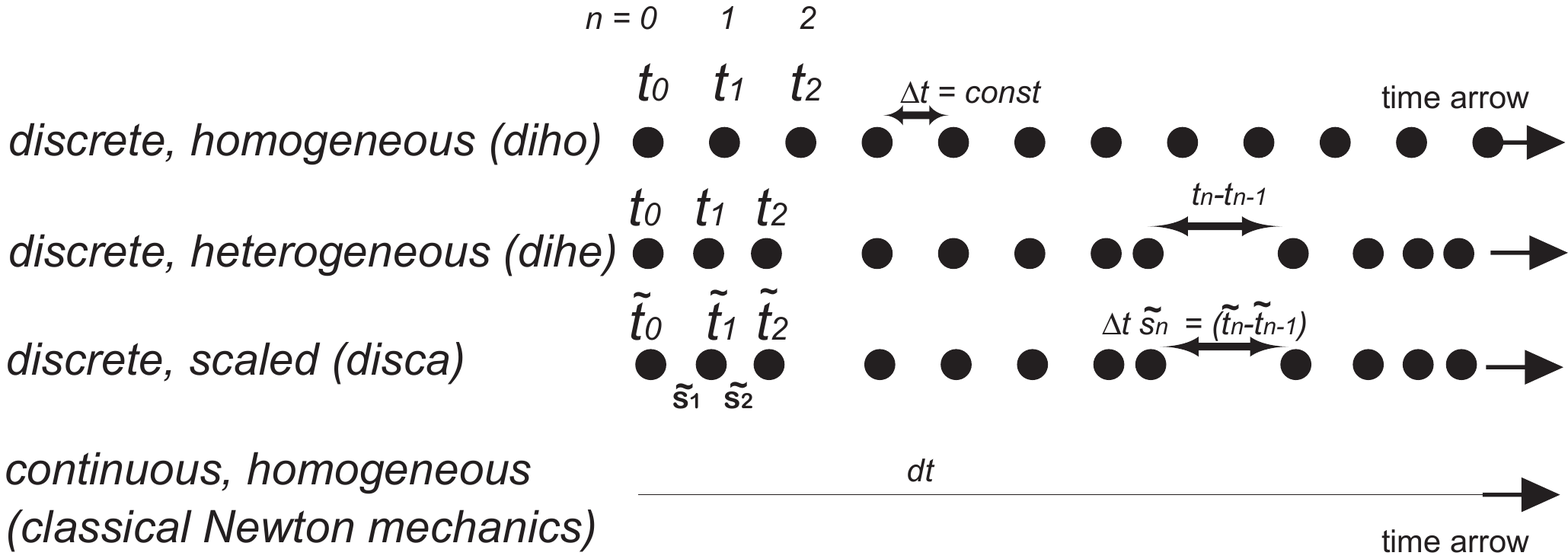}

Figure 2: The various discrete mechanics description used in this
manuscript are displayed side by side including the nomenclature used.
The discrete time points are shown and labeled in part. Also, the
time difference between time points are indicated. 

Indeed, by inspection of eq. 15 there is the possibility to obtain
also in presence of a discrete time and a complex force (i.e. $F(r_{n})\neq const$)
time reversibility by introducing time as a discrete dynamic variable
$\tilde{t}_{n}$. If time is a discrete dynamic variable it is a function
of $n$ and the step sizes $\tilde{t}_{n}-\tilde{t}_{n-1}$ are variable
as indicated in Figure 2. This so-called discrete scaled representation
(abbreviated by the disca representation in Figure 2; note this representation
has in addition a scaling variable $s_{n}$ to be introduced below)
enables to guarantee the time reversibility of the Newton law (eq.
14 ) by requesting\\
 
\begin{equation}
\boxed{\frac{\tilde{t}_{n+1}-\tilde{t}_{n}}{\tilde{t}_{n}-\tilde{t}_{n-1}}=\frac{F(r_{n})}{F(r_{n+1})}}
\end{equation}

Equation (16) is called in the following the reversibility axiom.
(Note, eq. 16 is not defined for $F(r_{n+1})=0$ unless $F(r_{n})$
is also 0 and thus with $F(r_{n})=const=0$ constant time steps are
obtained; correspondingly eq. 15 is fullfilled for $\frac{\partial V(r_{n})}{\partial r_{n}}=\frac{\partial V(r_{n+1})}{\partial r_{n+1}}$
= 0).

With other words, if the potential, respectively its corresponding
force, is of complex nature (i.e. $F(r_{n})\neq const$) time is a
dynamic variable following eq. 16 to guarantee time reversibility
of the system of interest. The request of a dynamic discrete time
if the potential is of complex nature has already been mentioned by
Lee (1983). One may argue now, that the discrete scaled time description
may be superior to the discrete homogeneous one because it enables
a time reversible description. However, the following caveats appear:
(i) The description of complex physical systems is reversible, which
is against our daily experience. Thus, by the introduction of time
reversibility consequently the arrow of time got lost. (ii) In a three
dimensional space, time has to be expanded to a tensor of second rank
as discussed in details in the next chapter. (iii) A further complexity
arises in a many particles system as discussed later yielding time
to be a multi-dimensional tensor. These arguments indicate again that
$\Delta t_{n}=const$ (possibly in the order or equal to the Planck
time). Under this assumption time reversibility of a system with a
complex potential is lost and hence an arrow of time is introduced
automatically without the need to introduce a macromolecular ensemble
description that follows thermodynamics. 

However, by establishing a discrete dynamic time variable combined
with eq. 16 a first important connection between the time irreversible
discrete, homogeneous mechanics (diho) and its acclaimed approximation
the Newton mechanics has been obtained by establishing a discrete
but time heterogenous description (disca) of mechanics (compare the
characteristics of the three systems in Figure 2). In the next chapters
the introduction of the time scaling variable $s_{n}$ will allow
to complete the connection between the two mechanics theories, yielding
a microscopic entropy description as we shall see. With other words,
it is the authors assumption in the following that the time quantum
is constant (and therefore independent of the step number $n$ or
the potential present) and the dynamic nature of the time is only
of use to calculate the progression of the energy as discussed below,
respectively, to guarantee energy conservation. Before we go on, we
make a short excursion into the description of time reversibility
in a three dimensional space.

\subsection{Time reversibility/irreversibility in a three dimensional space}

Let us in the following consider the evolution of the Newton's law
of a single particle in presence of a discrete time in a three dimensional
space under the request of time reversibility. By trying to expand
our finding about force-dependent scaling of time from the one dimensional
case to the corresponding three dimensions the complication arises
that the force is now a vector (i.e. $\mathbf{F_{n}}(\mathbf{r}_{n})=(F^{x}(\mathbf{r}_{n}),\, F^{y}(\mathbf{r}_{n}),\, F^{z}(\mathbf{r}_{n})$),
while so far time was a scalar. In order to enable time reversibility,
time is expanded to a diagonal tensor of second rank (e.g. a 2x2 matrix
with only diagonal elements):

\begin{equation}
t_{n}\rightarrow\left[\begin{array}{ccc}
t_{n}^{x} & 0 & 0\\
0 & t_{n}^{y} & 0\\
0 & 0 & t_{n}^{z}
\end{array}\right]=\mathbf{t}_{n}\mathbf{1}
\end{equation}
with $\mathbf{1}$ being the unity matrix and $\mathbf{t\mathrm{_{n}=\left[\begin{array}{c}
t_{n}^{x}\\
t_{n}^{y}\\
t_{n}^{z}
\end{array}\right]}}$

This extension should be regarded as a mathematical trick as mentioned
above in order to enable a formalism for time reversibility under
a discrete time. 

In accordance to the two step process in one dimension described above,
the corresponding two step process in a three dimensional space of
one step forward followed by a backward step is described by 
\begin{equation}
\dot{\mathbf{r}}_{n}=\dot{\mathbf{r}}_{n+2}=\dot{\mathbf{r}}_{n+1}+(\mathbf{t}_{n+1}-\mathbf{t}_{n})\mathbf{1}\mathbf{F}(\mathbf{r}_{n+1})=\dot{\mathbf{r}}_{n}-(\mathbf{t}_{n}-\mathbf{t}_{n-1})\mathbf{\mathbf{1}F}(\mathbf{r}_{n})+(\mathbf{t}_{n+1}-\mathbf{t}_{n})\mathbf{\mathbf{1}F}(\mathbf{r}_{n+1})
\end{equation}
The time reversibility of the Newton's law requests the time is scaled
by 
\begin{equation}
\frac{\tilde{t}{}_{n+1}^{x}-\tilde{t}{}_{n}^{x}}{\tilde{t}{}_{n}^{x}-\tilde{t}_{n-1}^{x}}=\frac{F^{x}(\mathbf{r}{}_{n})}{F^{x}(\mathbf{r}_{n+1})},\:\frac{\tilde{t}{}_{n+1}^{y}-\tilde{t}{}_{n}^{y}}{\tilde{t}{}_{n}^{y}-\tilde{t}{}_{n-1}^{y}}=\frac{F^{y}(\mathbf{r}{}_{n})}{F^{y}(\mathbf{r}_{n+1})},\:\frac{\tilde{t}{}_{n+1}^{z}-\tilde{t}{}_{n}^{z}}{\tilde{t}{}_{n}^{z}-\tilde{t}{}_{n-1}^{z}}=\frac{F^{z}(\mathbf{r}_{n})}{F^{z}(\mathbf{r}_{n+1})}
\end{equation}
Note, that while in each dimension there appears a separate time scaling
and this scaling is dependent on the corresponding spatial component
of the forces present, the time evolution of the system is still discrete
and the time evolution is defined by the number of steps $n$ and
not the step size. While this approach works mathematically it is
composed of a bizarre complexity (i.e. multi dimensional time) and
hence it is regarded only as a mathematical trick to get insights
into entropy as we shall see in the following.

\subsection{Scaling of time in accordance to Nosé}

For simplicity we go back to the single particle in a one dimensional
space and its description by the discrete constant time description
(diho in Figure 2). We now introduce the scaling variable $\tilde{s}{}_{n}$
and scale in parallel the time to yield 
\begin{equation}
\boxed{\tilde{s}{}_{n}(t_{n}-t_{n-1})\,=\tilde{t}_{n}-\tilde{t}_{n-1}}
\end{equation}
with $t_{n}-t_{n-1}=\Delta t\,=const$. (Please note, that the $s_{n}$
is defined as the corresponding variable $s$ of Nosé with $dt=\frac{d\tilde{t}}{\tilde{s}}$,
following the denotation for \textasciitilde{} used in the review
by Hünenberger (2005). In this new so called discrete, scaled representation
(so called disca representation) described in Figure 2 the time $ $$\tilde{t}_{n}\,$
is still a dynamic variable, while the system is extended by the scaling
variable $\tilde{s}{}_{n}.$ The transformation from the diho to the
disca representation is thus obtained by

\begin{equation}
\tilde{r}_{n}=r_{n}\,,\,\dot{\tilde{r}}_{n}=\frac{\dot{r}_{n}}{s_{n}}\,,\,\tilde{s}{}_{n}=s_{n}\,,\,\mathrm{and\:}\dot{\tilde{s}}{}_{n}=\frac{\dot{s}_{n}}{s_{n}}.
\end{equation}

$\dot{\tilde{s}}_{n}$ (and correspondingly $\dot{s}{}_{n}$) is thereby
defined by $\dot{\tilde{s}}_{n}=\frac{\tilde{s}_{n+1}-\tilde{s}_{n}}{\tilde{t}_{n}-\tilde{t}_{n-1}\,}$
defining it forward in time $\dot{\tilde{s}}_{n}(\tilde{t}_{n+1},\tilde{t}_{n},\tilde{t}_{n-1})$
as it has been done for $\ddot{\mathit{\mathrm{\mathit{r}}}}\mathit{_{n}}(t_{n+1},t_{n},t_{n-1})$
(see Figure 1). 

Such an approach of scaling of time has been introduced by Nosé. He
established the mathematical formulation of the so called Nosé-Hoover
thermostat (Nosé, 1984a and 1984b; Nosé, 1986; Hoover, 1985; Hoover,
2007; Martyna et al., 1992) enabling the molecular dynamics simulation
of a system to sample configurations of a canonical (constant-temperature)
ensemble. It is based on the extension of the real system by an artificial
dynamical time scaling variable $s$. This important relationship
serves in the following for the establishment of the microscopic entropy
of a single particle (note, that $s_{n}$ is the discrete analog of
$s$). Most importantly thereby is by doing so Nosé introduced a Lagrangian
that in presence of the dynamical variable $s$ yields in addition
to the Newton law, also a Hamiltonian and guarantees a Boltzmann weighted
canonical ensemble desription. Furthermore, its ensemble-averaged
Hamiltonian is thereby a constant of motion referring to the energy
of the system. Hence, the Nosé-Hoover thermostat introduces a Lagrangian
that describes a system on the microscopic level under scaling of
time, which yields upon ensemble averaging a macroscopic thermodynamic
description of the system bridging the microscopic with the macroscopic
world. 

Following Nosé and the transformation rules of Table 1, the discrete
time analog of the Nosé Lagrangian at time point $n$ for a single
particle is then given by
\begin{equation}
L_{n}^{N}=\frac{1}{2}\frac{m\left(\tilde{r_{n}}-\tilde{r}_{n-1}\right)^{2}\tilde{s}_{n}^{2}}{(\tilde{t}_{n}-\tilde{t}_{n-1})^{2}}-V(\tilde{r}_{n})+\frac{1}{2}Q\dot{\tilde{s}}_{n}^{2}-g\, k_{B\,}T\, ln\tilde{s}_{n}
\end{equation}
where $\, k_{B}$ is the Boltzmann constant, a constant Q, which has
been described as a ``mass''-like term for the motion of $\tilde{s}_{n}$
with Q > 0 with actual units of energy time squared, and g is equal
to the number of degrees of freedom in the real system $N_{df}$.(Note,
since $s_{n}$ is according to the reversibility axiom dependent on
the force $F$ as we shall see below the system of interest is not
an extended system as in the case of Nosé and therefore $g=N_{df}$
and not $N_{df}+1$ as it is the case for the Nosé extended system.
In the absence of a requested reversibility the system would be extended
though. Furthermore, we would like to mention that although under
the given conditions of having a single particle in a one dimensional
space g could by defined, the variable is kept here). The first two
terms of the Lagrangian represent the kinetic energy minus the potential
energy of the system, while the third and fourth terms represent the
kinetic energy minus the potential energy associated with the $\tilde{s}_{n}$
variable. The third term can thus be interpreted as a kinetic energy
of a heat bath coupled to the system of interest, and the fourth term
describes the heat transfer between the heat bath and the system of
interest. The heat bath mimics a bath composed of an infinite collection
of harmonic oscillators or one with an infinite number of ``particles''
in a box and thus shows an infinite heat capacity (Campisi and Hänggi,
2013). The heat or temperature bath is however of another nature than
usually defined in thermodynamics because it is a heat bath of the
unitless time scaling variable $\tilde{s}_{n}$ with a mass-like Q
with units $Js^{2}$ (energy{*}seconds{*}seconds) and a velocity $\dot{\tilde{s}}_{n}$
with unit $s^{-1}$. Nonetheless, it enables the use of a temperature
as highlighted by the explicit presence of the temperature $T$ in
the fourth term of the Lagrangian although the system described is
of microscopic nature. 

By applying the discrete Lagrangian equation (eq. 11) the Newton's
equation in the new representation is given by

\begin{equation}
\frac{1}{\tilde{t}_{n}-\tilde{t}_{n-1}}(\dot{\tilde{r}}_{n+1}-\dot{\tilde{r}}_{n}\frac{\tilde{s_{n}}^{2}}{\tilde{s}_{n+1}^{2}})=\frac{m^{-1}F(\tilde{r}_{n})}{{\tilde{s}_{n+1}^{2}}}
\end{equation}
which can be simplified to

\begin{equation}
\boxed{\ddot{\tilde{r}}_{n}=\frac{(\dot{\tilde{r}}_{n+1}-\dot{\tilde{r}}_{n})}{\tilde{t}_{n}-\tilde{t}_{n-1}}=\frac{m^{-1}F(\tilde{r}_{n})}{{\tilde{s}_{n+1}^{2}}}-\text{\ensuremath{\tilde{\lyxmathsym{\textgreek{g}}}}}_{n}\dot{\tilde{r}}_{n}}
\end{equation}

This Newton equation is extended by a friction term $ $$\text{\ensuremath{\tilde{\lyxmathsym{\textgreek{g}}}}}_{n}=\frac{\tilde{s}_{n+1}-\tilde{s}_{{n}}}{(\tilde{t}_{n}-\tilde{t}_{n-1})\:\tilde{s}_{n+1}}(1+\frac{\tilde{s}_{n}}{\tilde{s}_{n+1}})=\frac{\dot{\tilde{s}}_{n}}{\tilde{s}_{n+1}}(1+\frac{\tilde{s}_{n}}{\tilde{s}_{n+1}})$.
Hence, after transformation into the discrete, scaled so called disca
system (Figure 2) the Newton equation is extended by a friction term
which is attributed and dependent on the variables $\tilde{s}_{n}$
and $\tilde{s}_{n+1}$ as well as the corresponding time step. While
this equation is in principle reversible, it is interesting to note,
that it could be irreversible if the friction term is constrained
to values $\geqq0$. 

Equivalent to eq. 24 is the following equation 
\begin{equation}
\ddot{\tilde{r}}_{n}=\frac{m^{-1}F(\tilde{r}_{n})}{{\tilde{s}_{n}^{2}}}-\frac{\dot{\tilde{s}}_{n}}{\tilde{s}_{n}}(1+\frac{\tilde{s}_{n+1}}{\tilde{s}_{n}})\dot{\tilde{r}}_{n+1}
\end{equation}

It is now requested that the system in the disca frame must be reversible.
Hence, using the Newton equation a step forward yields

\begin{equation}
\dot{\tilde{r}}_{n+1}=\frac{m^{-1}F(\tilde{r}_{n})}{\tilde{s}_{n+1}^{2}}(\tilde{t}_{n}-\tilde{t}_{n-1})+\dot{\tilde{r}}_{n}\frac{\tilde{s}_{n}^{2}}{\tilde{s}_{n+1}^{2}}
\end{equation}

the second step backward is then 
\begin{equation}
\dot{\tilde{r}}_{n+2}=-\frac{m^{-1}F(\tilde{r}_{n+1})}{\tilde{s}_{n+2}^{2}}(\tilde{t}_{n+1}-\tilde{t}_{n})+\dot{\tilde{r}}_{n+1}\frac{\tilde{s}_{n+1}^{2}}{\tilde{s}_{n+2}^{2}}
\end{equation}

if time is reversible $\dot{\tilde{r}}_{n+2}=\dot{\tilde{r}}_{n}$
and $ $$\tilde{s}_{n+2}=\tilde{s}_{n}$ and thus

\begin{equation}
\dot{\tilde{r}}_{n}=-\frac{m^{-1}F(\tilde{r}_{n+1})}{\tilde{s}_{n+2}^{2}}(\tilde{t}_{n+1}-\tilde{t}_{n})+(\frac{m^{-1}F(\tilde{r}_{n})}{\tilde{s}_{n+1}^{2}}(\tilde{t}_{n}-\tilde{t}_{n-1})+\dot{\tilde{r}}_{n}\frac{\tilde{s}_{n}^{2}}{\tilde{s}_{n+1}^{2}})\frac{\tilde{s}_{n+1}^{2}}{\tilde{s}_{n+2}^{2}}
\end{equation}
\begin{equation}
\dot{\tilde{r}}_{n}=-\frac{m^{-1}F(\tilde{r}_{n+1})}{\tilde{s}_{n+2}^{2}}(\tilde{t}_{n+1}-\tilde{t}_{n})+\frac{m^{-1}F(\tilde{r}_{n})}{\tilde{s}_{n+2}^{2}}(\tilde{t}_{n}-\tilde{t}_{n-1})+\dot{\tilde{r}}_{n}
\end{equation}

yielding the reversibility axiom in the disca representation to be
\begin{equation}
\boxed{\frac{\tilde{s}_{n+1}}{\tilde{s}_{n}}=\frac{\tilde{t}_{n+1}-\tilde{t}_{n}}{\tilde{t}_{n}-\tilde{t}_{n-1}}=\frac{F(\tilde{r}_{n})}{F(\tilde{r}_{n+1})}}
\end{equation}

In the disca frame the system is thus reversible if equation (30)
is fulfilled.

\subsection{The microscopic entropy of a single particle}

Following Nosé (Hünenberger, 2005; Nosé 1984a) from the Lagrangian
(eq. 22) in the disca representation the corresponding Hamiltonian
can now be written 

\begin{equation}
H_{n}^{N}=\frac{1}{2}\, m\dot{\tilde{r}}_{n}^{2}\,\tilde{s}_{n}^{2}+V(\tilde{r}_{n})+\frac{1}{2}Q\dot{\tilde{s}}{}_{n}^{2}+gk_{B}T\, ln\,\tilde{s}_{n}\label{energy-1}
\end{equation}
This function is a constant of the motion and evaluates to the total
energy of the disca system. 

Nosé and Hoover figured that the equations of motion can be reformulated
to go back to a representation that has a homogenous time sampling
yielding the disco representation defined in Figure 2 (Nosé, 1986;
Hoover, 1985; Hünenberger, 2005). The transformation from the disca
system to the disco system variables is thereby achieved through 
\begin{equation}
r_{n}=\tilde{r}_{n}\,,\,\dot{r}_{n}=\dot{\tilde{r}}_{n}\tilde{s}{}_{n}\,,\,\tilde{s}{}_{n}=s_{n}\,,\, F(r_{n})=F(\tilde{r}_{n})\,,\,\dot{s}_{n}=\tilde{s}{}_{n}\dot{\tilde{s}}{}_{n}\,,\,
\end{equation}

\[
\mathrm{and\:}\mathbf{\ddot{\mathit{\mathrm{\mathit{r}}}}\mathit{_{n}}}=s_{n+1}s_{n}\ddot{\tilde{r}}_{n}+\dot{\tilde{r}}_{n}\dot{s}_{n}=s_{n+1}s_{n}\ddot{\tilde{r}}_{n}+\dot{\tilde{r}}_{n}\tilde{s}{}_{n}\dot{\tilde{s}}{}_{n}=s_{n}^{2}\ddot{\tilde{r}}_{n}+\dot{\tilde{r}}_{n+1}\tilde{s}{}_{n}\dot{\tilde{s}}{}_{n}\,
\]

Based on these expressions, the Lagrangian equations of motions can
be rewritten

\begin{equation}
\ddot{\mathit{\mathrm{\mathit{r}}}}\mathit{_{n}}=\frac{1}{m}F(r_{n})-\text{\textgreek{g}}_{n}\dot{r}_{n+1}
\end{equation}
with $\text{\textgreek{g}}_{n}=\dot{s}_{n}s_{n}^{-1}=\frac{s_{n+1}-s_{n}}{\text{\textgreek{D}t}\, s_{n}}=\frac{1}{\text{\textgreek{D}}t}(\frac{s_{n+1}}{s_{n}}-1)$ 

An equivalent expression is

\begin{equation}
\ddot{\mathit{\mathrm{\mathit{r}}}}\mathit{_{n}}=\frac{s_{n}}{s_{n+1\,}m}F(r_{n})-\frac{\dot{s}_{n}}{s_{n+1}}\dot{r}_{n}=\frac{1}{m}F(r_{n+1})-\frac{\dot{s}_{n}}{s_{n+1}}\dot{r}_{n}
\end{equation}

The constant of motion evaluating to the total energy of the entire,
pseudo extended system is given by
\begin{equation}
H_{n}^{N}=\frac{1}{2}\, m\dot{r}_{n}^{2}\,+V(r_{n})+\frac{1}{2}Q\frac{\dot{s}{}_{n}^{2}}{s_{n}^{2}}+gk_{B}T\, ln\, s_{n}
\end{equation}

(note, however, that this term is no longer a Hamiltonian; Hünenberger,
2005). 

This constant of motion is composed of the inner energy of the system
$U$ described by the first two terms followed by the energy of the
bath and the exchange energy (i.e. the latter two terms having the
variable $s_{n}$). The resemblance of this term with the free energy
(or Helmholtz Energy) $A\,=U-T\text{}S$ is next put forward (note,
the variable $A$ is used instead of the traditionally used letter
$F$ for the Helmholtz energy, because $F$ is used here for the force).
It is intriguing to define a microscopic entropy of a single particle
at time point $t_{n}$ to be

\begin{equation}
S_{n}=-g\, k_{B}\, ln\, s_{n}
\end{equation}

yielding 

\begin{equation}
H_{n}^{N}=\frac{1}{2}\, m\dot{r}_{n}^{2}\,+V(r_{n})+\frac{1}{2}Q\frac{\dot{s}{}_{n}^{2}}{s_{n}^{2}}-T\, S_{n}=U+\frac{1}{2}Q\frac{\dot{s}{}_{n}^{2}}{s_{n}^{2}}-T\, S_{n}
\end{equation}

By using in the definition the word ``microscopic'', we follow a
suggestion by Prigogine (1997) because $S_{n}$ is a non ensemble-averaged
term. Note, since we are in a time reversible description, the microscopic
entropy is in general of reversible character and thus not a monotonously
increasing quantity (see below and 3.11).

\subsection{The microscopic entropy of a system with many particles}

To extend the system from one particle to Z particles with Z very
large (we are still in a one dimensional space and the particles are
not interacting with each other keeping the potential simple $V_{i}(r_{i,n})$)
for each particle $i$ there is a separate scaling of time denoted
$s_{i,n}$ that follows the reversibility axiom $\frac{s_{_{i,n+1}}}{s_{i,\, n}}=\frac{F_{i}(r_{i,n})}{F_{i}(r_{i,n+1})}$
which yields
\begin{equation}
H_{n}=\sum_{i=1}^{Z}\frac{1}{2}\, m_{i}\dot{r}_{i,n}^{2}\,+V_{i}(r_{i,n})+\frac{1}{2}Q_{i}\frac{\dot{s}{}_{i,n}^{2}}{s_{i,n}^{2}}+g_{i}k_{B}T\, ln\, s_{i,n}
\end{equation}

and correspondingly, the microscopic entropy of a Z particles system
is defined as
\begin{equation}
S_{n}=-\sum_{i=1}^{Z}g_{i}\, k_{B\,}\, ln\, s_{i,n}
\end{equation}
In an attempt to simplify eq. 39 by reducing amongst others the number
of scaling factors $s_{i,n}$ the following description of the microscopic
entropy is introduced by assuming there are $J$ $ $groups of particles
and within each group the particles have undistinguishable properties
but are still distinct since the system is microscopically described
(i.e. within a group each particle has $g_{j}$ degrees of freedom,
the same responds to the potential, the same scaling factor, and the
same velocity and mass). We define $\, p_{j}=\frac{\#\, of\, particles\, in\, group\, j}{Z}=\frac{Z_{j}}{Z}$
with $\sum_{j=1}^{J}p_{j\,}=1$. For very large systems (i.e. at the
thermodynamic limit) $p_{j}$ is also the probability of a particle
to be in the $j$ group. Thus, the derived microscopic entropy of
the Z particles system at time point $t_{n}$ is given by
\begin{equation}
S_{n}=-Z\sum_{j=1}^{J}p_{j\,}g_{j}\, k_{B\,}\, ln\, s_{j,n}\,=<-k_{B}\, g\, ln\, s_{n}>=-k_{B}\, N_{df}\,<ln\, s_{n}>
\end{equation}

with $g=Z\, g_{i}$. In the latter part of eq. 40 it is assumed that
all particles have the same degree of freedom and thus $g=Z\, g_{i}\,=N_{df}$
with $N_{df}$ being the degree of freedom of the entire system. 

In the next step, it is assumed that the scaling of time is very small
(i.e. $s_{j,n}$ is very close to 1). This assumption called in the
following the ``slow changing force limit'' is valid if the change
of the force from one time step to the next is very small (i.e. $F(r_{n+1})\approx F(r_{n})+\text{\textgreek{D}}$
with $\text{\textgreek{D}}$ small, see also eq. 30). Support for
a $s_{j,n}$ close to 1 is also based on the notion that scaling of
time is usually not directly observed in physics. Under this assumption
the $ln$ of eq. 40 can be described by a Taylor expansion of first
order and the averaging can be put inside the $ln$, which results
in

\begin{equation}
S_{n}=-k_{B\,}N_{df}\,<ln\, s_{n}>\approx-k_{B\,}N_{df}\,\sum_{j=1}^{J}p_{j\,}\,(s_{j,n}-1)=-k_{B\,}N_{df}\,[(\sum_{j=1}^{J}p_{j\,}\, s_{j,n})-1]
\end{equation}

\begin{equation}
S_{n}\approx-k_{B\,}N_{df}\, ln\sum_{j=1}^{J}p_{j\,}\, s_{j,n}=-k_{B\,}N_{df}\,\, ln<s_{n}>_{\,}
\end{equation}

In accordance, the microscopic entropy difference between two time
points $n$ and $m$ for a $Z$ particles system with having all particles
the same $g_{i}$ is given by

\begin{equation}
\text{\textgreek{D}}S\,=\, S_{m}-S_{n}=-k_{B}\, N_{df}(<ln\, s_{m}>-<ln\, s_{n}>)\approx-k_{B\,}N_{df}\, ln\,\frac{<s_{m}>}{<s_{n}>}
\end{equation}

It is interesting to note that this description can be used to calculate
for a Z particles system the microscopic entropy difference between
two time points by using averaged time scaling factors. 

Furthermore, in contrast to the macroscopic thermodynamic entropy,
the microscopic entropy can also be calculated for a system with a
single particle or a few particles at any given time point $n$. In
addition, the microscopic entropy difference can be calculated without
having the system in both states in equilibrium and the averaging
is not over all possible states as in statistical mechanics, but over
the observed state. However, the change of the microscopic entropy
is in general not a monotonously increasing quantity as requested
for the macroscopic entropy, an issue to be discussed below (3.11).
Nonetheless, with the above mathematical trickery a close resemblance
between the microscopic and macroscopic entropy is obtained.

\subsection{Boltzmann entropy versus the microscopic entropy of a many particles
system}

It is attempted in the following to show a profound relationship between
the microscopic entropy and the corresponding macroscopic one. Let
us start with the Boltzmann entropy which shall be defined as the
macroscopic entropy of a large system of Z independent, identical,
indistinguishable particles (such as the ideal gas or diluted gas)
for which each micro-state has the same probability. Boltzmann showed
that the entropy is then given by

\begin{equation}
S^{B}=-k_{B}\,<ln\, p_{j}>=-k_{B}Z\sum_{j=1}^{J}p_{j\,}ln\, p_{j}
\end{equation}
with $p_{j}=\frac{Z_{j}}{Z}$ being the probability of a single particle
to be in the state $j$ and the average is a single particle average
taken over all the possible states $J$ of the particle. The derivation
of this formulation of the Boltzmann entropy is starting with the
well known formula

\begin{equation}
S^{B}=k_{B}\, ln\, W
\end{equation}
with 
\begin{equation}
W=\frac{Z!}{\prod_{j=1}^{J}Z_{j}!}
\end{equation}
describing the number of micro-states that the Z particles system
can adopt if there are $J$ distinguishable groups of states that
each is composed of $Z_{j}$ particles (Greiner et al., 1993).

Using the Stirling formula (i.e. $ln\, Z!=Z\, ln\, Z\,-Z$ or $Z!=(\frac{Z}{e})^{Z}$
) which is true for large $Z$, eq. 45 can be simplified to

\begin{equation}
S^{B}=k_{B}\,(ln\, Z!\,-{\displaystyle \sum_{j=1}^{J}\, lnZ_{j}!)}=k_{B}(Z\, lnZ\,-Z-[\sum_{j=1}^{J}Z_{j}\, lnZ_{j}-Z_{j}])
\end{equation}

\[
=k_{B}(Z\, lnZ\,-Z-\sum_{j=1}^{J}Z_{j}\, ln\, Z_{j}+\sum_{j=1}^{J}Z_{j})=k_{B}(Z\, lnZ\,-\sum_{j=1}^{J}Z_{j}\, ln\, Z_{j})
\]
\begin{equation}
=k_{B}\, Z\,(ln\, Z-\sum_{j=1}^{J}p_{j}\, ln\,[p_{j}Z])=k_{B}\, Z\,(ln\, Z-\sum_{j=1}^{J}p_{j}\, ln\, p_{j}-\sum_{j=1}^{J}p_{j}\, ln\, Z)=-k_{B}Z\sum_{j=1}^{J}p_{j}\, ln\, p_{j}
\end{equation}

In comparison, the microscopic entropy of this Z particles system
is given by $S_{n}=-k_{B}\, Z\,\sum_{j=1}^{J}p_{j}\, ln\, s_{j,n}$
with $g_{j}=1$, which is for example the case for an ideal monoatomic
gas in a one dimensional volume. Because the description is microscopic
each particle is distinct and thus requests its own scaling of time
(i.e. $s_{j,n}$). In the ``slow exchanging force limit'' the averaging
can be put inside the $ln$ (eq. 42) 

\begin{equation}
S_{n}\approx-\, k_{B\,}Z\, ln<s_{n}>=-k_{B\,}Z\, ln\,\sum_{j=1}^{J}p_{j\,}s_{j,n\,}
\end{equation}

Since the macroscopic entropy is only defined by an averaging, which
in the case of interest can be done by a single particle averaging
(eqs. 44-48), only the single particle average of the microscopic
entropy can be compared with its macroscopic counterpart yielding
with $\text{\textgreek{d}}_{lj}$ being the Kronecker's delta function
\begin{equation}
<S_{n}>\approx-k_{B\,}Z<ln\,\sum_{j=1}^{J}p_{j\,}s_{j,n\,}>=-k_{B\,}Z\sum_{l=1}^{J}\, p_{l}\,\text{\textgreek{d}}_{lj}\, ln(\sum_{j=1}^{J}p_{j}\, s_{j,n})=-k_{B\,}Z\sum_{l=1}^{J}\, p_{l}\, ln\,(p_{l}\, s_{l,n})
\end{equation}

A classical thermodynamic system in equilibrium for which the entropy
is calculated can be assumed to be in the ``slow changing force''
limit with $s_{l,n}$ close to 1, yielding

\begin{equation}
<S_{n}>\approx-k_{B\,}Z\sum_{l=1}^{J}p_{l\,\,}ln(p_{l\,}s_{l,n})\approx-k_{B\,}Z\sum_{l=1}^{J}p_{l\,\,}ln\, p_{l\,}=S^{B}
\end{equation}

Thus, for a system in the ``slow changing force limit'' with independent
non interacting particles, the average microscopic entropy in equilibrium
approximates the Boltzmann entropy of the system. An interpretation
of this analogy suggests that the original description of the entropy
by Boltzmann is just a statistical averaging of a variable which is
close to 1 (i.e. $s_{l,n}$) but that the entropy term itself has
not its root in statistical mechanics but rather in the discreteness
of time.

\subsection{Gibbs entropy versus the microscopic entropy of a many non-interacting
particles system}

Above, we have shown that the average microscopic entropy at equilibrium
of a system of Z independent, non-interacting, indistinguishable particles
approximates to the Boltzmann entropy.$ $ However, in more complex
systems the above defined Boltzmann entropy leads to increasingly
wrong predictions of entropies and physical behaviors, because it
ignores the interactions and correlations between different particles.
Instead one must follow Gibbs, and must consider the ensemble of states
of the system as a whole, rather than single particle states. If there
are totally $\text{\textgreek{W}}$ micro-states $k$ that all together
describe the same macro-state the following description corresponds
to the Gibbs entropy

\begin{equation}
S^{G}=-k_{B}\sum_{k=1}^{\lyxmathsym{\textgreek{W}}}p_{k\,}ln\, p_{k}
\end{equation}

with $p_{k}$ being the probability of a macro-state to be in the
micro-state $k$ with $\sum_{k=1}^{\lyxmathsym{\textgreek{W}}}p_{k\,}=1$
(note, $p_{k}$ is thus defined differently than above).

In a micro canonically behaving system the probability of each micro-state
is equal with $p_{k}=\frac{1}{\lyxmathsym{\textgreek{W}}}$ yielding
the famous entropy formula (Greiner et al., 1993) 
\begin{equation}
S^{G}=k_{B}\, ln\,\text{\textgreek{W}}
\end{equation}

In comparison, the microscopic entropy of a Z non-interacting particles
system was described by $S_{n}=-\, k_{B\,}Z\,<ln\, s_{n}>$ with $<ln\, s_{n}>$
the sum of $ln\, s_{n,i}$ over all particles $i$ multiplied by $\frac{1}{Z}$,
which is equal to the average over the particles in the thermodynamic
limit ($Z\rightarrow\text{\ensuremath{\infty}}$ and $V\rightarrow\text{\ensuremath{\infty}}$).
In addition, in the thermodynamic limit, the ensemble averaging of
a function over all the micro-states of the macro-state is identical
to the averaging over the particles if they do not interact with each
other yielding $Z\,<ln\, s_{n}>_{particle\, averaging}\,=<ln\, s_{n}>_{ensemble\, averaging}=\sum_{k=1}^{\text{\textgreek{W}}}(p_{k\,}ln\, s_{k,n})\approx ln\sum_{k=1}^{\text{\textgreek{W}}}(p_{k\,}\, s_{k,n})$
following the ideas of eqs. 41-43. Thus,

\begin{equation}
S_{n}\approx-k_{B\,}\, ln\sum_{k=1}^{\text{\textgreek{W}}}(p_{k\,}\, s_{k,n})
\end{equation}

Since the macroscopic entropy is only defined for an ensemble, only
the average over all the micro-states of the microscopic entropy can
be compared with its macroscopic counterpart yielding 
\begin{equation}
<S_{n}>\approx-k_{B\,}<ln\sum_{k=1}^{\text{\textgreek{W}}}(p_{k\,}\, s_{k,n})>=-k_{B\,}\sum_{m=1}^{\lyxmathsym{\textgreek{W}}}\, p_{m}\,\text{\textgreek{d}}_{mk}[ln\sum_{k=1}^{\text{\textgreek{W}}}(p_{k\,}\, s_{k,n})]=-k_{B\,}\sum_{m=1}^{\text{\textgreek{W}}}\, p_{m}\, ln(p_{m\,}\, s_{m,n})
\end{equation}

The delta function $\text{\textgreek{d}}_{lj}$ thereby guarantees
the independency of each micro-state.

Assuming the system to be in the ``slow changing force'' limit the
ensemble-average microscopic entropy in equilibrium is given by

\begin{equation}
<S_{n}>\approx-k_{B\,}\sum_{m=1}^{\text{\textgreek{W}}}p_{m}\, ln\,(p_{m\,})=S^{G}
\end{equation}

Thus, for a many non interacting particles system in equilibrium in
the ``slow changing force'' limit, the microscopic entropy averaged
over all the possible micro-states approximates the Gibbs entropy
of the system. The Gibbs and Boltzmann entropies are thus derived
from a particle-individual scaling of time. The individual time scalings
were thereby requested by the hypothesis of a discreteness of time
in combination with the acclaimed time reversibility nature of the
established microscopic laws of physics. 

It is interesting to note however, that individual time scaling in
presence of a continuous time would also yield a microscopic entropy,
which would correspond to the Gibbs entropy if ensemble averaged.
While the author prefers scaling of time to be a consequence of the
simple and sound rational of time being discrete, its continuos analog
may have also its merits.

\subsection{Gibbs entropy versus the microscopic entropy of a system with interacting
particles}

It remains to be shown that the ensemble average of the microscopic
entropy approximates the Gibbs entropy also in a system with many
($Z$) particles which may interact with each other. In order to describe
such systems, a few concepts have been established in statistical
mechanics of which we will discuss here the mean field approximation
and the ``two particles only'' or pair approximation (Greiner et
al., 1993). 

In the presence of an intermolecular potential $V(r_{i,n},r_{k,n})$
with a long range in distance such as gravitation with its $\frac{1}{|r_{i,n}-r_{k,n}|}$
dependency, it appears that the individual contributions of the interactions
$\frac{1}{2}{\displaystyle \sum_{i,k=1(i\neq k)}^{Z}V(}r_{i,n},r_{k,n})$
in the Hamiltonian/Lagrangian can be approximated by a mean potential
described by $\frac{1}{2}V(r_{i,n})$. This description can be interpreted
as the potential that all the particles together induce at the position
$r_{i}$ at the given time point $n$. Obviously, with this approximation
the particles are statistically independent of each other relaxing
the acclaimed complexity that arises from switching the description
from a particle to an ensemble point of view, and thus ensemble averaging
and single particle averaging are equivalent. Furthermore, since the
mean potential is now of the form of the potential introduced above
(for example eqs. 10 and 38) the above derivation of the macroscopic
Gibbs entropy from the microscopic one (eq. 56) can be applied also
to interacting particles if their interaction is described adequately
with a mean field approximation.

While for long range potentials the mean field approximation appears
to be reasonable, it fails for short range potentials such as the
interatomic potential in a diluted real gas. In such systems the pair
approximation can be put forward, which assumes that the system is
composed of two particles only. For diluted systems under short range
potentials this approximation is reasonable, because for most of the
states not more than two particles will be present within the relevant
action distance of the short range potential at any given time yielding
a system of statistically independent pairs of particles. The pair
of particles is thus the smallest unit for averaging and pairs of
particles are independent of each other yielding an equivalence between
ensemble averaging over the pair and single pair averaging. If the
pair potential acts onto the two particles equivalently with $|F(r_{i,n},\, r_{k,n})|=|F(r_{k,n},\, r_{i,n})|$
the two particles show also the same scaling factor $s_{i,n}=s_{k,n}$
(because of the reversibility axiom) and thus the microscopic entropy
of the pair is given by $S_{n}=-\, k_{B\,}2\, ln\, s_{n}$ yielding
for the microscopic entropy of the Z particles system $S_{n}\approx-\, k_{B\,}\frac{Z}{2}\,2\, ln\,<s_{n}>_{pair\, average}=-\, k_{B\,}\, ln\,<s_{n}>_{ensemble\, average}$
and finally this results in $<S_{n}>\approx S^{G}$ upon ensemble
averaging of the microscopic entropy. 

It remains however to be shown that without any assumption of a model
the ensemble average of the microscopic entropy is equivalent to its
macroscopic analog.

\subsection{Example: The change of the entropy of a volume-expanding ideal gas}

Although it was demonstrated above theoretically that the ensemble-averaged
microscopic entropy approximates the Gibbs entropy, an example is
given next for furhter illustration. By doing it, we will however
elucidate an important not yet in details considered problem of the
theory introduced, which is the problem of time-irreversibility in
time reversible descriptions of physical laws (chapter 3.11). 

In the example the entropy change of the expansion of an ideal gas
is calculated following the microscopic entropy expression derived
above. The isothermal reversible expansion (i.e. $T=const$) is along
$r$ of an ideal gas located in a box with the area A and the side
$r$ from the initial Volume $V_{i}=r_{i\,}A$ to the final volume
$V_{f}=r_{f}\, A$. From a macroscopic view the expansion of the Z
particles system is based on a force $F$ which acts on the area $A$
giving rise to the pressure $p=\frac{F}{A}$ . Using the conventional
thermodynamics the change of the macroscopic entropy is calculated
to be $\Delta S^{G}=S_{f}-S_{i}=Z\, k_{B}\,\ln\frac{V_{f}}{V_{i}}$. 

Using the microscopic entropy definition of the Z particles system
(eq. 43) the change of the microscopic entropy of an expanding gas
can be described by

\begin{equation}
\text{\textgreek{D}}S\,=\, S_{f}-S_{i}\approx-k_{B\,}Z\, ln\,\frac{<s_{f}>}{<s_{i}>}=-k_{B\,}Z\, ln\,\frac{<F(r_{i})>}{<F(r_{f})>}=-k_{B\,}Z\, ln\,\frac{F_{i}}{F_{f}}
\end{equation}

For this derivation, the reversibility axiom of eq. (16) was used
(note, it has not been used above for the derivation of the ensemble-averaged
microscopic entropy). Furthermore, it was assumed that the averaged
force $<F(r_{n})>$ that acts on the particle is equal to the macroscopic
force $F_{n}=p_{n\,}A$ (this relationship is based on the principle
that in a time reversible isothermal process the split of a force
into a sum of smaller forces must follow the summation rule because
of the impossibility to make a perpetum mobile).

Furthermore, during the isothermal expansion of the gas the following
dependence between the force and the volume is given through the ideal
gas equation $pV=Z\, k_{B\,}T$ (note, the ideal gas equation is believed
to be correct also under a discrete time, because of its time insensitive
character and the experimental verification). 
\begin{eqnarray}
dpV+pdV=0\\
dF\frac{V}{A}+\frac{F}{A}dV=0\\
\int\frac{dF}{F}= & -\int & \frac{dV}{V}\\
\Rightarrow & ln\frac{F_{f}}{F_{i}} & =-ln\frac{V_{f}}{V_{i}}
\end{eqnarray}
For the microscopic entropy change we finally get
\begin{equation}
\Delta S=S_{f}-S_{i}=-Z\, k_{B}\ln\frac{V_{f}}{V_{i}}
\end{equation}
This obtained entropy change is equal to the macroscopic entropy change
$\Delta S^{G}$ multiplied by minus 1 using the classical description
of entropy. Obviously, the factor -1 is disturbing. An attempt to
resolve this apparent discrepancy is given in the next chapter.

\subsection{The problem of time-irreversibility in time reversible descriptions
of physical laws}

It has been mentioned that there is a fundamental problem to connect
the reversible physical description of a microscopic system with its
macroscopic analog usually described by statistical mechanics, because
it appears to be difficult or impossible to connect a reversible description
with an irreversible one (Prigogine, 1997; Prize, 1996). In the approach
presented, this problem is still puzzling as demonstrated with the
example above on the expansion of the idea gas and as discussed in
the following. 

We did start with a time-irreversible microscopic description of a
system under a complex potential by introducing the discreteness of
time and by defining the time steps to be of constant size (i.e. $\Delta t_{n}=const$).
However, since most of the physical laws are time reversible, we (re)introduced
time reversibility including the time reversibility axiom (i.e. eq.
16: $\frac{s_{n+1}}{s_{n}}=\frac{t_{n+1}-t_{n}}{t_{n}-t_{n-1}}=\frac{F(r_{n})}{F(r_{n+1})}$)
by scaling of time yielding a time reversible description of the energy
of the system including the constant of motion evaluating to the entire
energy with the microscopic entropy (i.e. $H_{n}=\frac{1}{2}\, m\dot{r}_{n}^{2}\,+V(r_{n})+\frac{1}{2}Q\frac{\dot{s}{}_{n}^{2}}{s_{n}^{2}}+gk_{B}T\, ln\, s_{n}$
and $S_{n}=-g\, k_{B}\, ln\, s_{n}$ for a single particle, eqs. 37-40).
Thus, both the microscopic entropy as well as the reversibility axiom
are reversible. With other words, for the reversibility axiom and
the energy description with the microscopic entropy the arrow of time
is lost. The consequence of the loss of the arrow of time for the
reversibility axiom is that it is not possible to distinguish between
a forward and a backward process. Mathematically, this means that
the reversibility axiom is either the one defined in eq. 16: $\frac{\tilde{s}_{n+1}}{\tilde{s}_{n}}=\frac{F(\tilde{r}_{n})}{F(\tilde{r}_{n+1})}$
or $\frac{\tilde{s}_{n+1}}{\tilde{s}_{n}}=\frac{F(\tilde{r}_{n+1})}{F(\tilde{r}_{n})}$
derived from the corresponding backward process starting with $\dot{r}_{n-1}=\dot{r}_{n}-\frac{1}{m}(t_{n+1}-t_{n})\frac{\partial V(r_{n})}{\partial r_{n}}$
in analogy to eqs. 12-15 (i.e. $\dot{r}_{n-2}=\dot{r}_{n-1}-\frac{1}{m}(t_{n}-t_{n-1})\frac{\partial V(r_{n-1})}{\partial r_{n-1}}$
followed by the reversion of time of the second step $\dot{r}_{n-2}=\dot{r}_{n-1}+\frac{1}{m}(t_{n}-t_{n-1})\frac{\partial V(r_{n-1})}{\partial r_{n-1}}=\dot{r}_{n}-\frac{1}{m}(t_{n+1}-t_{n})\frac{\partial V(r_{n})}{\partial r_{n}}+\frac{1}{m}(t_{n}-t_{n-1})\frac{\partial V(r_{n-1})}{\partial r_{n-1}}=\dot{r}_{n}$
and $\frac{\partial V(r_{n-1})}{\partial r_{n-1}}=\frac{\partial V(r_{n+1})}{\partial r_{n+1}}$).
Hence, the loss of the arrow of time due to the introduction of a
time reversible discrete physics introduces the following ambiguity
for the reversibility axiom: 

\begin{equation}
\,\frac{\tilde{s}_{n+1}}{\tilde{s}_{n}}=\frac{F(\tilde{r}_{n})}{F(\tilde{r}_{n+1})}\, or\,\:\frac{\tilde{s}_{n+1}}{\tilde{s}_{n}}=\frac{F(\tilde{r}_{n+1})}{F(\tilde{r}_{n})}
\end{equation}

This finding resolves apparent problems such as the expanding gas
discussed above (chapter 3.10). Indeed, in the case of the expanding
ideal gas, the introduction of the absolute value removes the negative
sign in the calculation of the macroscopic entropy change from its
microscopic analog yielding an approximate equivalency between the
ensemble-averaged microscopic entropy and its macroscopic counterpart. 

The remaining caveat is that the constant of motion $H_{n}=\frac{1}{2}\, m\dot{r}_{n}^{2}\,+V(r_{n})+\frac{1}{2}Q\frac{\dot{s}{}_{n}^{2}}{s_{n}^{2}}-T\, S_{n}$
with $S_{n}=-g\, k_{B}\, ln\, s_{n}$ is still time reversible. If
however, the ``mass'' $Q$ of the heat bath (or $Q_{i}$ of all
the water baths) goes towards infinity ($Q\rightarrow\infty,\, Q_{i}\rightarrow\infty$)
the oscillatory nature of the Nosé-Hoover ``thermostat'' is lost
and concomitantly time irreversibility is obtained. Note, in molecular
dynamics simulations, the value of $Q$ is chosen such, that during
a calculation the temperature fluctuates around the temperature of
the temperature bath. If $Q$ is set small, there is a fast oscillation,
if it is too long the oscillation is too slow and very long simulations
are required to obtain a canonical distribution of the system. In
the case of $Q\rightarrow\infty$, $s=1$ and thus a regular molecular
dynamics simulation without thermostating is established (Nosé, 1984;
Nosé 1986; Hoover, 1985; Hunenberger, 2005). In the discrete time
description of the $H_{n}$ however $s_{n}\neq1$ (unless $F(r_{n})=const$)
and thus time irreversibility is obtained. 

This ad hoc introduction of requesting $Q\rightarrow\infty$ in order
to introduce a time irreversible description of the system is strengthened
by the following argument. Let us again consider the time reversible
evolution of a single particle under a given potential $V(r_{n})$
at time point $n$ described by the Nosé-Hoover Lagrangian (eq. 31;
Hünenberger, 2005; Nosé 1984a) 
\[
\]
\begin{equation}
L_{n}^{N}=\frac{1}{2}\, m\dot{r}_{n}^{2}\,-V(r_{n})+\frac{1}{2}Q\frac{\dot{s}{}_{n}^{2}}{s_{n}^{2}}-gk_{B}T\, ln\, s_{n}
\end{equation}
The time reversible nature of the velocity has been demonstrated above
resulting in the reversibility axiom (chapter 3.4). However, also
the evolution of the scaling factor $s_{n}$ must be time reversible.
With the definition of the scaling factor velocity $\dot{s}_{n}=\frac{s_{n+1}-s_{n}}{t_{n}-t_{n-1}\,}$
(see chapter 3.4) the scaling factor evolves in a first step forward
to 

\begin{equation}
s_{n+1}=s_{n}+\dot{s}_{n}(t_{n}-t_{n-1})
\end{equation}
\[
\]
and accordingly for the second step

\begin{equation}
s_{n+2}=s_{n+1}+\dot{s}_{n+1}(t_{n+1}-t_{n})
\end{equation}
If the second step is now backward in time

\begin{equation}
s_{n+2}=s_{n+1}-\dot{s}_{n+1}(t_{n+1}-t_{n})=s_{n}+\dot{s}_{n}(t_{n}-t_{n-1})-\dot{s}_{n+1}(t_{n+1}-t_{n})
\end{equation}
With the request of time reversibility $s_{n}=s_{n+2}$ which simplifies
the above equation by using $t_{n}-t_{n-1}=const$ to 

$ $
\begin{equation}
\dot{s}_{n}=\dot{s}_{n+1}
\end{equation}
$ $which is true for a two step process with a step forward in time
followed by a step backward in time. This result is obvious since
the change of scaling of time in the step $n\rightarrow n+1$ should
be equal to the change of scaling of time of the step backward $n+1\rightarrow n$
multiplied by $-1$. 

In addition, also the scaling factor $s_{n}$ must follow (its) (discrete)
Lagrangian equation (i.e. $\frac{\text{}1}{t_{n}-t_{n-1}}(\frac{\partial L(s_{\mathrm{n+1}}\mathrm{,\mathrm{\mathbf{\dot{s}}_{\mathrm{n+1}})}}}{\mathbf{\partial\dot{s}_{\mathrm{\mathrm{n+1}}}}}-\frac{\partial L(\mathbf{s}_{\mathrm{n}}\mathrm{,\mathrm{\mathbf{\dot{s}}_{\mathrm{n}})}}}{\partial\mathbf{\dot{s}_{\mathrm{\mathrm{n}}}}})=-\frac{\partial L}{\partial\mathbf{s}_{n}}$)
yielding

\begin{equation}
\frac{\text{}1}{t_{n}-t_{n-1}}(Q\frac{\dot{s}{}_{n+1}}{s_{n+1}^{2}}-Q\frac{\dot{s}{}_{n}}{s_{n}^{2}})=\frac{3}{2}Q\frac{\dot{s}{}_{n}^{2}}{s_{n}^{3}}+gk_{B}T\frac{1}{s_{n}}
\end{equation}
\begin{equation}
\frac{\text{}Q}{t_{n}-t_{n-1}}(\frac{\dot{s}{}_{n+1}-\dot{s}{}_{n}+\dot{s}{}_{n}}{1}-\dot{s}{}_{n}\frac{s_{n+1}^{2}}{s_{n}^{2}})=(\frac{3}{2}Q\frac{\dot{s}{}_{n}^{2}}{s_{n}^{3}}+gk_{B}T\frac{1}{s_{n}})s_{n+1}^{2}
\end{equation}

\begin{equation}
Q(\frac{\dot{s}{}_{n+1}-\dot{s}{}_{n}}{t_{n}-t_{n-1}})=Q\,\ddot{\mathbf{\mathrm{s}}}_{n}=(\frac{3}{2}Q\frac{\dot{s}{}_{n}^{2}}{s_{n}^{3}}+gk_{B}T\frac{1}{s_{n}})\, s_{n+1}^{2}-\dot{s}{}_{n}(1-\frac{s_{n+1}^{2}}{s_{n}^{2}})
\end{equation}

Since the right hand of eq. 76 is not evidently 0, but with the above
eq. 73 (i.e. $\dot{s}{}_{n+1}-\dot{s}{}_{n}=0$) the left hand is
0 for $Q\neq\infty$ , eq. 76 can only be fulfilled with a $Q\rightarrow\infty$.
Thus, $Q\rightarrow\infty$ is a requirement and a consequence of
the request of a time reversible scaling factor yielding thereby a
time irreversible constant of motion referring to the total energy
as expected from a thermodynamic point of view. With setting the ``mass''
$Q$ of the water bath (or $Q_{i}$ of all the water baths in the
case of a multi particle system) towards infinity ($Q\rightarrow\infty$),
the ``kinetic'' energy of the water bath of the scaling variable
$s_{n}$ (or $s_{i,n}$) is infinite and thus has an infinite capacity
to take up the energy that is lost in the discrete time steps of the
system. This description may open a discussion about the nature of
the water bath (baths), which is not a regular thermodynamic water
bath because it deals with the unitless scale variable $s_{n}$ ($s_{i,n}$)
outside the space-time. While on the one hand it is possible to interpret
the scaling of time as an extension of the space-time by additional
dimensions (i.e. for each particle three additional dimensions $s_{i,x,y\, or\, z}$)
as has been done by Nosé in the Nosé-Hoover thermostat (however with
a single $s$ only), the author prefers the view, that the introduction
of the scaling of time is just a mathematical construct that enables
to preserve the first law of thermodynamics, which describes energy
conservation.

\subsection{Time progression of a discrete system step by step}

For a simulation of a time irreversible system comprising one particle
(or many particles) it is potentially interesting to write down the
progression of the system step by step. Having constant time steps
$\Delta t$ the evolution of a single particle system in a one dimensional
space starting under the (boundary) conditions

$V(r_{n}),F(r_{n}),\, m,\,\dot{r}_{1},\, t_{1},\, r_{1},$ and $\Delta t$
can be calculated step by step as follows:

Using the (discrete) Newton equation 

$\textrm{\ensuremath{\ddot{\mathbf{\textrm{r}}}_{1}}=}\frac{1}{m}F(r_{1})$

Using the definition of the discrete $\textrm{\ensuremath{\ddot{\mathbf{\textrm{r}}}_{n}}}$
the velocity of step 2 can be obtained

$\dot{r}_{2}=\frac{1}{m}F(r_{1})\,\text{\textgreek{D}}t+\dot{r}_{1}$

with the definition of the velocity the coordinate at step 2 can then
be obtained

$r_{2}=\dot{r}_{2}\text{\textgreek{D}}t+r_{1}$

which allows the calculation of the acceleration by

$\textrm{\ensuremath{\ddot{\mathbf{\textrm{r}}}_{2}}=}\frac{1}{m}F(r_{2})$

The corresponding microscopic entropy change is then given by

$\Delta S=S_{2}-S_{1}=k_{B}(|ln\frac{F(r_{1})}{F(r_{2})}|)$

Correspondingly, the step $n+1$ can be derived from values from step
$n$ by

$\dot{r}_{n+1}=\frac{1}{m}F(r_{n})\,\text{\textgreek{D}}t+\dot{r}_{n}$

$r_{n+1}=\dot{r}_{n+1}\text{\textgreek{D}}t+r_{n}$

$\textrm{\ensuremath{\ddot{\mathbf{\textrm{r}}}_{n+1}}=}\frac{1}{m}F(r_{n+1})$
and

$\Delta S=S_{n+1}-S_{n}=k_{B}(|ln\frac{F(r_{n})}{F(r_{n+1})}|)$.

\section{Conclusion}

It was demonstrated that under the hypothesis that time is discrete
with constant time steps the microscopic physical laws in presence
of a complex force ($F(r)\neq const)$ are time irreversible. Upon
the introduction of the scaling of the discrete time in order to reintroduce
time reversibility a microscopic entropy is obtained, which ensemble
average in equilibrium under the ``slow changing force'' limit appoximates
the macroscopic Boltzmann/Gibbs entropy. Thus, an alternative microscopic
description of entropy is derived. Since this approach does not rely
on the statistical mechanics approach of Gibbs and Boltzmann, the
Loschmidt and Zermelo\textquoteright{}s reversal and recurrence objections
appear to be resolved (Steckline, 1982; Lohschmidt, 1876; Poincare,
1890), and an arrow of time is obtained already at the microscopic
level of physics. 

The now arising question is, which entropy derivation is the correct
one, the established Gibbs and Boltzmann ensemble-averaged macroscopic
entropy or the presented one with the microscopic entropy? In favor
of the presented theory are the reversal and recurrence objections
and the establishment of a microscopic arrow of time that is connected
to the second law of thermodynamics. However, the continuos time descriptions
of physics introduced by Newton is highly successful and thus has
obviously also its merits. In a first attempt towards answering the
question raised, it is required to highlight the differences between
the two entropy descriptions (eqs. 53 and 56):

\[
S^{G}=-k_{B}\sum_{k=1}^{\text{\textgreek{W}}}p_{k\,}ln\, p_{k}
\]

and

\[
<S_{n}>\approx-k_{B\,}\sum_{m=1}^{\text{\textgreek{W}}}\, p_{m}\, ln(p_{m\,}\, s_{m,n})
\]

Thus, the ensemble averaged microscopic entropy approximates its macroscopic
counterpart only at the ``slow changing force'' limit at which $s_{m,n}$
is close to 1.

With other words, only at the ``slow changing force'' limit the
two entropy descriptions are equivalent. However, if the acting force
changes considerably within a time step $\text{\textgreek{D}}t$,
which has been hypothesized to be the Planck time (i.e. $5.4\,10^{-44}$
s), the ensemble averaged microscopic entropy is distinct from its
Gibbs analog. If experimental data under such conditions would be
available insight into the nature of the entropy might be possible.
Such conditions may have been present at the beginning of the universe
during which the acting potential must have been changed very fast.
In this context it is interesting to mention that the observed inflation
of the universe might support the presented theory, because the inflation
can apparently be explained by a scaling of time (Masreliez, 2004;
Penrose, 2010; Rovelli, 2011; Smolin, 2013), and the scaling of time
is at the heart of the presented theory being a consequence of a time-reversible
description of an irreversible process under a discrete time. In return,
if time is discrete in nature, the apparent inflation is an artifact
of the established time reversible description of a time irreversible
universe. Another area of potential interest, that could eventually
help in the elucidation of the origin of the entropy, might be the
analysis of the observed time asymmetry of the weak force in particle
physics (Christenson et al., 1964). 

In conclusion, an alternative microscopic derivation of entropy is
given, which originates from the quantization of time, which is a
concept that has only occasionally been investigated in the past.
For the author this theory resolves the many objections that have
been raised for the Boltzmann/Gibbs entropy and introduces an arrow
of time at the microscopic physics level. However, whether time is
discrete and concomitantly the presented microscopic entropy is an
adequate microscopic representation of the macroscopic entropy state
function remains to be demonstrated (experimentally).

\section{Acknowledgment}

We would like to thank extensively Dr. Dominik Leitz, Prof. Dr. Gunnar
Jeschke, Prof. Dr. Klaus Hepp, Dr. David Neuhaus, Prof. Dr. Martin
Quack, Prof. Dr. Markus Reiher, Prof. Dr. Hans-Jakob Wörner, Dr. Julien
Orts, Daniel Blattmann and Prof. Roger Penrose for critical and helpful
discussions. This work we dedicate to Prof. Dr. Martin Quack on teh
occasion of his status emeritus.

\author{}

\section{References}

L. Boltzmann (1866). Über die Mechanische Bedeutung des Zweiten Hauptsatzes
der Wärmetheorie. Wiener Berichte 53: 195\textendash{}220

P. Caldorola (1953) A new model of classical electron. Supplmento
al Nuovo Cimento 10: 1747-1804.

M. Campisi, P. Hänggi (2013) Thermostated Hamiltonian Dynamics with
Logs Oszillators. J. Phys Chem. B. online.

J. H. Christenson, J. W. Cronin, V. L. Fitch, R. Turlay(1964) Evidence
for the 2\textgreek{p} Decay of the K20 Meson. Phys. Rev. Letters
13, 138-140.

R. Clausius (1865). The Mechanical Theory of Heat \textendash{} with
its Applications to the Steam Engine and to Physical Properties of
Bodies. London: John van Voorst.

J.A. Cadzow (1970) Discrete calculus of variations. Inst. J. Control
11, 393-407.

R.A.H. Farias, E. Recami (2007) Introduction of a Quantum of Time
(``chorion'') and its Concequences for Quantum Mechanics. arXiv:quant-ph/9706059

W. Greiner, L. Neise, H. Stöcke (1993) Thermodynamik und Statistische
Mechnik. Verlag Harri Deutsch, Frankfurt am Main.

W.G. Hoover (1985) Canonical dynamics: Equilibrium phase-space distributions.
Phys. Rev. A 31, 1695-1697.

W.G. Hoover (1999) Time reversibility, computer simulation, and chaos.
Adv. Series in nonlinear dynamics 13, 1-262 (ISBN 981-02-4073-2)

W.G. Hoover (2007) Nose-Hoover nonequilibrium dynamics and statistical
mechanics. Mol. Simulation 33, 13-19.

P.H. Hünenberger (2005) Thermostat Algorithms for Molecular Dynamics
Simulations. Adv. Polym. Sci 173, 105-149.

G. Jaroszkiewicz, K. Norton (1997) Principles of discrete time mechanis:
I. Particle systems. J. Phys. A.: Math. Gen 30, 3115-3144.

G. Jaroszkiewicz, K. Norton (1997) Principles of discrete time mechanis:
I. Classical field theory. J. Phys. A.: Math. Gen 30, 3145-3164.

G. Jaroszkiewicz, K. Norton (1998) Principles of discrete time mechanis:
III. Quantum field theory. J. Phys. A.: Math. Gen 31, 977.1000

T.D. Lee (1983) Can time be a discrete dynamical variable? Physics
Letters 122B, 217-220.

Robert Levi (1927) Theorie de l'action universelle et discontunue.
Journal de Physique et le Radium 8, 182-198.

L.D. Landau, E.M. Lifshitz (1980). Statistical Physics. 5 (3 ed.).
Oxford: Pergamon Press. ISBN 0-7506-3372-7. Translated by J.B. Sykes
and M.J. Kearsley

J.P. Lees et al. (2012) Observation of Time-Reversal Violation in
the B0 Meson System. Phys. Review Letters. 109, 211801.

J. Loschmidt, (1876) Sitzungsber. Kais. Akad. Wiss. Wien, Math. Naturwiss.
Classe 73, 128\textendash{}142. 

G.J. Martyna, M.L. Klein, M. Tuckerman (1992) Nose-Hoover hains: The
canonical ensemble via contnuous dynamics. J. Chem. Phys. 97, 2635-2643.

C.J. Masreliez (2004) Sclae Expanding Cosmos: Theory I - An introduction.
Apeiron 11, 99-133.

S. Nosé (1984) A unified formulation of the constant temperature molecular
dynamics methods. J. Chem. Phys. 81, 511-519.

S. Nosé (1984) A molecular dynamics method for simulations in the
canonical ensemble. Mol. Phys. 52, 255-268.

S. Nosé (1986) An extension of the canonical ensemble molecular dynamics
method. Mol. Phys. 57, 187-191.

H. Prize (1996) Time's arrow and Archimedes' Point: New directions
for the physics of time. ISBN 0-19-510095-6

R. Penrose (1989) The emperor's new mind. Oxfort University Press.
ISBN 0-19-851973-7

R. Penrose (2010) Cycles of Time. The Bodley Head, London. ISBN 9780224080361

H. Poinca\'{r}e (1890) Sur le probz`eme des trois corps et les\'{ }equations
de la dynamique, Acta Math. 13, 1\textendash{}270.

H. Poinca\'{r}e (Derniere Pensees (Paris, 1913)

I. Prigogine (1997) The End of Certainty: Time, Chaos and the New
Laws of Nature. The Free Press, New York.

C. Rovelli (2011). Zakopane lectures on loop gravity. 1102. pp. 3660.
arXiv:1102.3660. Bibcode 2011arXiv1102.3660R

L. Smolin (2013) Time Reborn. Houghton Micclin Harcourt, Boston 2013.
ISBN 978-0-547-51172-6.

H.S. Snyder (1947) Quantisized Space Time. Phys. Rev. 71, 38.

V.S. Steckline (1982) Zermelo, Boltzmann, and the recurrence paradox.
Am. J. Phys. 51 894-897.

J.J. Thomson (1925) The intermittence of eletric force, Proc. Roy.
Soc. of Edinburgh, 46, 90.

M.C. Valsakumar (2005) Stochasticity, decoherence and an arrow of
time from the discretization of time? Pramana 64, 593-606. 

W. van Gunsteren, X. Daura, A.E. Marc (2002) Computation of free energies.
Helvetica Chimica Acta. 85, 3113-3129.

C.N. Yang (1947) On quantized space-time. Physical Review 72, 874
\end{document}